\documentclass{article}
\usepackage{spconf,amsmath,graphicx,hyperref}
\usepackage[utf8x]{inputenc}
\usepackage{textcomp}

\usepackage{booktabs}
\usepackage{transparent}
\usepackage{siunitx}
\usepackage{cleveref}
\usepackage{standalone}
\usepackage{tikz}
\usepackage{pgfplots}
\pgfplotsset{compat=1.17}
\usepackage{siunitx}
\usepackage{csquotes}
\usepackage{contour}
\contourlength{1pt}
\usetikzlibrary{
  positioning,
  arrows,
  arrows.meta,
  tikzmark,
  matrix,
  fit,
  calc,
  patterns,
  chains,
  scopes,
  shapes.multipart,
  decorations,
  decorations.markings,
  decorations.pathreplacing,
  backgrounds,
  shadows.blur,
  calligraphy,
  shadings,
  fadings,
  intersections,
  svg.path,
}

\ninept
\usepackage{amsmath}
\usepackage{amssymb}
\usepackage{bbm}
\usepackage{physics}
\usepackage{siunitx}
\usepackage{trfsigns}

\newcommand{\vect}[1]{\ensuremath{\boldsymbol{\mathbf{#1}}}}

\newcommand\speechsignal{\vect{s}}
\newcommand\hidden{\vect{h}}
\newcommand\hiddenseq{\vect{H}}
\newcommand\enc{\operatorname{Embed}}
\newcommand\dec{\operatorname{Dec}}
\newcommand\framescore{\ensuremath{q}}
\newcommand\uttscore{\ensuremath{y}}
\newcommand\featext{\operatorname{FeatExt}}
\newcommand\transformer{\operatorname{Enc}}
\newcommand\latent{\vect{z}}
\newcommand\latentseq{\vect{Z}}

\title{%
    SPEECH QUALITY-BASED LOCALIZATION OF LOW-QUALITY SPEECH AND TEXT-TO-SPEECH SYNTHESIS ARTEFACTS
}
\name{Michael Kuhlmann\thanks{Computational resources were provided by the Paderborn Center for Parallel Computing.}, Alexander Werning, Thilo von Neumann, Reinhold Haeb-Umbach}
\address{%
    Paderborn University, Germany
}
\begin{document}
\maketitle
\begin{abstract}
A large number of works view the automatic assessment of speech from an utterance- or system-level perspective. While such approaches are good in judging overall quality, they cannot adequately explain why a certain score was assigned to an utterance. frame-level scores can provide better interpretability, but models predicting them are harder to tune and regularize since no strong targets are available during training. In this work, we show that utterance-level speech quality predictors can be regularized with a segment-based consistency constraint which notably reduces frame-level stochasticity. We then demonstrate two applications involving frame-level scores: The partial spoof scenario and the detection of synthesis artefacts in two state-of-the-art text-to-speech systems. For the latter, we perform listening tests and confirm that listeners rate segments to be of poor quality more often in the set defined by low frame-level scores than in a random control set.
\end{abstract}

\section{INTRODUCTION}
Automatic speech quality assessment (SQA) models summarize the quality of a speech signal into a single scalar value.
This is useful for a handful of tasks like data cleaning, measuring the performance of voice transmission systems~\cite{mittag21_interspeech} or the evaluation of generative speech technology like text-to-speech (TTS)~\cite{huang22f_interspeech}.
Many of these predictors learn a mapping from the speech signal to a mean opinion score (MOS) by using training pairs that were queried during large-scale, often crowd-sourced, listening tests, asking participants to rate different constructs like overall quality~\cite{mittag21_interspeech} or naturalness~\cite{Cooper2021How}.
While these approaches do provide a picture of overall quality, they cannot explain what decision rules the predictor learns to quantify the quality as good or bad, or, more specifically, reveal something about the \emph{internal mapping} from smaller segments of speech to quality scores.


There exists some studies that developed subjective listening tests to \emph{localize} and \emph{identify} error patterns in TTS systems.
Edlund et al.~\cite{Edlund2015Audience} designed an audio response system that required participants to click a button every time noticed an error in the synthesis.
They had a professional speech synthesis developer classify the most often clicked locations as easily identifiable problems or not, but did not provide any results on the types of errors.
Gutierrez et al.~\cite{gutierrez21_ssw} used Rapid Prosody Transcription to identify prosody errors in TTS synthesis.
Participants were asked to mark words where the prosody did not sound natural and subsequently asked them to classify the error by choosing from four options.
While they could show that number and type of errors differ by task and system, the agreement among participants on the error locations was low.
Pine et al.~\cite{pine25_ssw} let a group of expert evaluators discuss on the existence, severity and type of errors by choosing from a list of different error types.

In a similar spirit to automating SQA, we would like to automate the \emph{localization} of speech synthesis errors.
As there exists no large-scale datasets for this task to train a model in a supervised fashion, the only way is to learn from utterance-level targets.
Quality-Net~\cite{Fu2018QualityNet} first looked at how to train SQA models to predict frame-level scores.
Kuhlmann et al.~\cite{kuhlmann25_interspeech} extended this idea to SQA models using pretrained self-supervised learned (SSL) encoders~\cite{huang25g_interspeech}.
While they could show that frame-level scores can be used to detect unnatural distortions, localized distortions also lowered clean frame-level scores in their vicinity, worsening detection precision.
In this work, we show that frame-level predictions of SQA models can be effectively regularized by using a consistency constraint~\cite{liu2024analyzing}.
We validate the effect of this regularization by measuring the detection performance of partially spoofed speech~\cite{Zhang2022ThePD}, where we can increase the precision of the detections from $31.7\%$ to $62.3\%$ by training with the consistency constraint.
Finally, we use the regularized frame-level scores to detect synthesis artefacts in two state-of-the-art TTS systems: StyleTTS2~\cite{li2023styletts} and F5-TTS~\cite{chen2024f5}.
We perform a listening test to 1) check whether this detection method does not produce an excessive number of false positives and 2) to classify the kind of artefacts the SQA model is sensitive to.
Compared to randomly extracted control samples, the segments detected by our method were considerably more often marked as detrimental.

\section{FRAME-LEVEL SPEECH QUALITY SCORES}
SQA models provide a prediction of speech quality for the full utterance.
The idea of frame-level scores is to provide an estimate of the speech quality at a finer resolution~\cite{Fu2018QualityNet,kuhlmann25_interspeech}.
As the name suggests, this is done at the finest resolution of a \emph{frame} which spans a few milliseconds of audio (e.g., \SI{25}{ms}).
We consider simple, but strong encoder-decoder SQA models~\cite{huang25g_interspeech} from which we extract frame-level scores.
Given a speech signal $\speechsignal=(s_1,\dots,s_N)$, the encoder extracts a frame sequence $\hiddenseq=\enc(\speechsignal)=(\hidden_1,\dots,\hidden_T)$ of embeddings of lower resolution ($T<N$).
The subsequent decoder can either process the time-pooled~\cite{manocha2021noresqa,ragano2024scoreq,tjandra2025aes} or frame-level embeddings and do the time-pooling afterwards~\cite{Cooper2021Generalization,mittag21_interspeech,Saeki2022UTMOS,wardah25_interspeech}, where only from the latter frame-level scores can be extracted~\cite{kuhlmann25_interspeech}.
Denote by $\vect{\framescore}=\dec(\hiddenseq)=(\framescore_1,\dots,\framescore_T)$ the sequence of frame-level scores and the utterance-level prediction by $\hat{\uttscore}=\operatorname{TimePool}(\vect{\framescore})$.
The SQA objective in its simplest form minimizes the error between predicted and annotated MOS $\uttscore$:
\begin{equation}
    \mathcal{L}_{\textrm{SQA}} = \lvert\hat{\uttscore}-\uttscore\rvert
\end{equation}
Kuhlmann et al.~\cite{kuhlmann25_interspeech} showed that frame-level scores obtained in this way could be used to detect artificial distortions in clean speech signals.
However, as only utterance-level targets are provided, there is no incentive for local embeddings to encode only local information, i.e., there persists a strong coupling between local and global scores~\cite{Fu2018QualityNet,kuhlmann25_interspeech}:
frame-level scores in the vicinity of a local artefact are smeared, making precise detections more difficult.

\subsection{Consistent frame-level scores}
To improve the detection precision of local artefacts, we require frame-level scores to be robust to changes in long context segments.
Fu et al.~\cite{Fu2018QualityNet} constrained the context information of a BLSTM encoder by initializing the forget gate bias with a large negative value and adding a frame-level constraint to the output scores:
\begin{equation}
    \mathcal{L} = \mathcal{L}_{\textrm{SQA}} + \alpha(\hat{\uttscore})\sum_{t=1}^T \left(\hat{\uttscore}-\framescore_{t}\right)^2,
\end{equation}
where $\alpha(\hat{\uttscore})=10^{\hat{\uttscore}-\uttscore_{\textrm{max}}}$ is a weighting factor that depends on the difference between the estimated quality and the maximum value of the quality scale $\uttscore_{\textrm{max}}$.
They observed that this constraint could significantly reduce the variance of frame scores obtained from clean speech.
However, the effect of the constraint exponentially decreases for lower scored signals.
For Transformer-based encoders, Kuhlmann et al.~\cite{kuhlmann25_interspeech} proposed to chunk the speech signal into blocks
of a few 100 milliseconds
and let the encoder process each block independently.
While this could successfully reduce the influence of long contexts, the detection performance dropped drastically because the model could no longer effectively utilize the full context.

In the training of neural audio codec models, Liu et al.~\cite{liu2024analyzing} noticed similar inconsistencies for the predicted tokens when an audio segment was embedded in context or not.
To improve consistency, i.e., an audio segment produces the same sequence of embeddings whether embedded in context or not, they cut and encode a random segment of the audio and minimize the error between the embeddings obtained from the full-context and from the slice.
We apply this idea to improve the consistency of frame-level scores of speech quality predictors.
Denote an SSL-based encoder $\enc$ by its feature extractor $\featext$ and Transformer encoder $\transformer$: $\enc=\transformer\circ\featext$.
The feature extractor yields latents $\latentseq=\featext(\speechsignal)$ of the same resolution as $\hiddenseq$, but encoding only limited local context.
During training, we sample a contiguous slice $\latentseq^{\textrm{slice}}=(\latent_m,\dots,\latent_M)$ with $1 \le m < M \le T$ from $\latentseq$ and encode it using the Transformer to obtain $\hiddenseq^{\textrm{slice}}=\transformer(\latentseq^{\textrm{slice}})$.
We extract the corresponding segment from $\hiddenseq$ and add the mean-squared error between the two embedding sequences to the total loss:
\begin{equation}
    \label{eq:emb-consistency}
    \mathcal{L} = \mathcal{L}_\textrm{SQA} + \lambda_{\textrm{emb}}\mathcal{L}_{\textrm{emb}}; \,\mathcal{L}_{\textrm{emb}}=\frac{1}{M-m}\sum_{t=m}^M\lVert\hidden_t-\hidden^{\textrm{slice}}_t\rVert_2^2,
\end{equation}
where $\lambda_{\textrm{emb}}$ is a weighting factor to balance the losses.

If a non-linear decoder with long-range context is used in the SQA model (e.g., a BLSTM as in UTMOS~\cite{Saeki2022UTMOS}), constraining the consistency of the embeddings alone may not be enough to reduce long-context dependence.
Therefore, we additionally add a consistency term to the frame-level scores.
Denote by $\vect{\framescore}^{\textrm{slice}}=\dec(\hiddenseq^{\textrm{slice}})$ the sequence of frame-level scores that are obtained by applying the decoder only to the embeddings obtained from the sliced latents $\latentseq^{\textrm{slice}}$.
The frame scores consistency constraint is then similar to the embedding consistency constraint~\labelcref{eq:emb-consistency}, but operating on the frame scores and using the mean absolute error:
\begin{equation}
    \label{eq:scores-consistency}
    \mathcal{L} = \mathcal{L}_\textrm{SQA} + \lambda_{\textrm{emb}}\mathcal{L}_{\textrm{emb}} + \lambda_{\textrm{scores}} \frac{1}{M-m}\sum_{t=m}^M\lvert\framescore_t-\framescore^{\textrm{slice}}_t\rvert.
\end{equation}

\section{APPLICATION TO PARTIAL SPOOF DETECTION}
To show that the consistency loss can
improve the localization of artefacts in speech, we consider the partial spoof scenario~\cite{Zhang2022ThePD}.
In this scenario, synthetic speech samples were substituted into bona fide speech and the task is to detect the edited regions.
In contrast to partial spoof countermeasures that are trained with knowledge about the edited regions, we train SQA models in the conventional way using human-annotated utterance-level scores only and apply a threshold to the continuous frame scores to obtain binary decisions.

\subsection{Implementation}
For training our models, we loosely follow the setup
in SHEET~\cite{huang25g_interspeech}
with some modifications\footnote{\url{https://github.com/fgnt/local_sqa}}: 1) We always use a WavLM~\cite{chen2022wavlm} model as encoder; 2) we perform $L_2$-normalization of the embedding sequence $\hiddenseq$ with $\lVert\bar{\hidden}\rVert_2$, where $\bar{\hidden}=\operatorname{TimePool}(\hiddenseq)$, which keeps the losses low at initialization; 3) we always include a contrastive loss with margin $0.1$~\cite{Saeki2022UTMOS} in $\mathcal{L}_{\textrm{SQA}}$; 4) we experiment with a 1-layer BLSTM decoder with 128 features per direction, followed by a linear projection layer.
\Cref{tab:ps-model-configurations} shows an overview of the model configurations with are tested on PartialSpoof~\cite{Zhang2022ThePD}.
We train all models on the combination of the main train track of the BVCC corpus~\cite{Cooper2021How} and the simulated train split (\texttt{TRAIN\_SIM}) of the NISQA corpus~\cite{mittag21_interspeech}.
All models first perform a loudness equalization of the input signal to $\SI{-18}{dBFS}$, followed by utterance-level mean and standard normalization.
We train all models for $\SI{100}{epochs}$ with an initial learning rate of $\SI{1e-4}{}$ which is linearly decayed to $\SI{1e-6}{}$.
For the consistency loss, we randomly cut slices between $\SI{200}{ms}$ and $\SI{1}{s}$.

\begin{table}[t]
    \centering
    \setlength{\tabcolsep}{11.5pt}
    \vspace{-6pt}
    \caption{%
        Tested model configurations for PartialSpoof.
        All models were trained on BVCC~\cite{Cooper2021How} + NISQA~\cite{mittag21_interspeech}.\strut
    }
    \begin{tabular}{lcccc}
        \toprule[1.5pt]
        \# & Encoder & Decoder & $\lambda_{\textrm{emb}}$ & $\lambda_{\textrm{scores}}$ \\
        \midrule[1pt]
        1 & WavLM Base & Linear & 0 & 0 \\ 
        2 & WavLM Base & BLSTM & 0 & 0 \\ 
        \midrule
        3 & WavLM Base & Linear & 1 & 0 \\ 
        4 & WavLM Base & BLSTM & 1 & 0 \\ 
        5 & WavLM Base & BLSTM & 0 & 1 \\ 
        6 & WavLM Base & BLSTM & 1 & 1 \\ 
        7 & WavLM Base & BLSTM & 10 & 1 \\ 
        8 & WavLM Large & BLSTM & 10 & 1 \\ 
        \bottomrule[1.5pt]
    \end{tabular}
    \label{tab:ps-model-configurations}
\end{table}

\subsection{BVCC results}
\Cref{tab:bvcc-test} shows results on the BVCC test split.
To show the effect of the consistency constraints, we report the mean \emph{frame score volatility}
\begin{equation}
    \sigma_{\vect{r}(\vect{\framescore})} = \sqrt{\frac{T}{\textrm{FPS}}}\operatorname{std}(\vect{r}(\vect{\framescore})),
\end{equation}
where $\textrm{FPS}$ is the number of frames per second and $\vect{r}(\vect{\framescore})=(r_1,\dots,r_{T-1})$ is the sequence of log-returns $r_t=\log\left(\frac{\framescore_{t+1}}{\framescore_t}\right)$.
The normalization factor changes the unit of time from a frame to a second.
A lower volatility means that the change in frame scores is less erratic between two frames, which may be helpful for interpretation.
Indeed, from the results, we find that the consistency losses can drastically reduce the volatility without negatively affecting the global quality estimates.
Note that a similar observation was made for Quality-Net~\cite{Fu2018QualityNet}.

\begin{table}[t]
    \centering
    \setlength{\tabcolsep}{12pt}
    \vspace{-6pt}
    \caption{%
        Results on BVCC test main.
        SRCC: Spearman rank correlation coefficient (higher is better).%
        \strut
    }
    \begin{tabular}{lcccc}
        \toprule[1.5pt]
         \# & Utterance SRCC & System SRCC & Volatility \\
         \midrule[1pt]
         1 & .864 & .904 & .510 \\
         2 & .862 & .904 & .355 \\
         \midrule
         3 & .862 & .905 & .172 \\
         4 & .868 & .915 & .099 \\
         5 & .865 & .915 & .061 \\
         6 & .870 & .917 & .064 \\
         7 & .871 & .922 & .055 \\
         8 & \textbf{.883} & \textbf{.923} & .091 \\
         \bottomrule[1.5pt]
    \end{tabular}
    \label{tab:bvcc-test}
\end{table}

\subsection{PartialSpoof results}
We apply the trained models to the PartialSpoof~\cite{Zhang2022ThePD} database without any fine-tuning.
The localization performance for models trained on PartialSpoof is usually evaluated using the \emph{segment-level} equal error rate and F1-score.
An alternative to segment-based evaluation is \emph{intersection-based} evaluation~\cite{Bilen2020Framework} which is more flexible by defining a \emph{detection tolerance} and a \emph{ground truth intersection} criterion.
The detection tolerance criterion counts a detected segment as \emph{relevant} if the intersection of the segment with any ground truth events amounts to at least the detection tolerance parameter $\rho_{\textrm{DTC}}$.
In a second step, the ground truth intersection criterion counts a ground truth event as \emph{true positive} if its intersection with any of the relevant detections amounts to at least the ground tolerance parameter $\rho_{\textrm{GTC}}$.

We measure detection performance with the intersection-based criterion with two configurations: $\rho_1=(\rho_{\textrm{DTC}}=0.7,\rho_{\textrm{GTC}}=0.3)$ and $\rho_2=(0.7,0.5)$.
Both configurations punish imprecise detections, but allow the model to only capture a part of a ground truth event (such as short synthesis artefacts in the substitutions).
As topline, we include BAM~\cite{zhong24_interspeech} which uses information about the spoofed locations during training and has state-of-the-art performance on PartialSpoof.
We extract frame-level predictions at a resolution of $\SI{20}{ms}$ from the pretrained checkpoint\footnote{\url{https://github.com/media-sec-lab/BAM}}, which matches the frame-level resolution of our SQA models.
We tune\footnote{\url{https://github.com/fgnt/sed_scores_eval}} the detection threshold for all models on the development split and report precision, recall and F1-score on the evaluation split.

\begin{table}[t]
    \centering
    \setlength{\tabcolsep}{4.5pt}
    \vspace{-6pt}
    \caption{%
        Detection results on the evaluation split of PartialSpoof.
        We use the intersection-based detection criterion with $\rho_1=(0.7,0.3)$ and $\rho_2=(0.7,0.5)$.
        All thresholds were tuned on the dev split.\strut
    }
    \begin{tabular}{lcccc}
        \toprule[1.5pt]
        \# & Volatility & Prec. ($\rho_1$/$\rho_2$) & Rec. ($\rho_1$/$\rho_2$) & F1 ($\rho_1$/$\rho_2$) \\
        \midrule[1pt]
        \transparent{0.5}{BAM~\cite{zhong24_interspeech}} & \transparent{0.5}{2.89} & \transparent{0.5}{.691/.651} & \transparent{0.5}{.762/.506} & \transparent{0.5}{.725/.569} \\
        \midrule
        1 & .376 & .263/.209 & \textbf{.545}/.446 & .355/.284 \\
        2 & .291 & .335/.258 & .523/\textbf{.467} & .408/.333 \\
        \midrule
        3 & .144 & .322/.233 & .395/.418 & .355/.300 \\
        4 & .083 & .461/.358 & .383/.349 & .419/.353 \\
        5 & .052 & .595/.508 & .339/.303 & .431/.379 \\
        6 & .057 & .584/.526 & .362/.309 & .447/.389 \\
        7 & .051 & .623/\textbf{.557} & .332/.296 & .434/.386 \\
        8 & .098 & \textbf{.628}/.520 & .404/.364 & \textbf{.492}/\textbf{.429} \\
        \bottomrule[1.5pt]
    \end{tabular}
    \label{tab:ps-detection}
\end{table}

The results are shown in \Cref{tab:ps-detection}.
The high recall of models $1$ and $2$ comes at the price of a high number of false positives, as indicated by the lowest precisions and F-scores among all models.
We observe that a reduction in volatility positively correlates with an increase in precision, but causes a decrease in recall performance.
Note that the partial spoof scenario is not a perfect match for our use case as many substitutions are of high quality.
Therefore, we expect to miss many spoof locations when using frame-level scores for detection (low \emph{recall}).
The high volatility of the BAM scores is due to the binary targets the model was trained with: A frame is either classified as bona fide (1) or spoofed (0), which promotes an erratic behavior of the frame predictions before the sigmoid activation.

\section{LOCALIZING SPEECH SYNTHESIS ARTEFACTS}
We now turn to the task of detecting and analyzing artefacts in synthesized speech samples.
As frame quality predictor, we use model configuration $8$ which achieved the best results on all tested datasets.

\subsection{Analyzed data}
We choose two state-of-the-art systems that showed to produce very human-like sounding speech~\cite{srinivasavaradhan25_interspeech}: F5-TTS~\cite{chen2024f5} and StyleTTS2~\cite{li2023styletts}.
Since the evaluation of TTS systems is very sensitive to the context and its intended application~\cite{edlund24_interspeech}, care must be taken when analyzing and giving meaning to artefact types.
As an application, we here opt for speech synthesis for audiobooks to showcase the procedure, potentials and pitfalls of this kind of analysis.
Our target data is the \texttt{test-clean} split of LibriTTS~\cite{zen19_interspeech} which we would like to clone as faithfully and artefact-free as possible\footnote{Despite its name LibriTTS contains read speech, not synthesized speech.}.
To test how successfully this can be done by F5 and StyleTTS2, we generate two speech samples from every transcription in \texttt{test-clean} using random references from the same for voice cloning, resulting in about $10,000$ speech samples to analyze for each TTS system.

\subsection{Localizing artefacts}
To localize artefacts, we apply a threshold to the frame-level scores.
Frames with quality scores below this threshold are deemed to be of detrimental quality and further analyzed as to what kind of artefact they contain (or whether they turn out to be a false positive).
To select a suitable threshold, we first collect frame-level scores from human speech in the domain of the target application, i.e., from \texttt{test-clean} in our case.
We then choose the threshold such that it satisfies a target false alarm rate for these scores, e.g., only $1\%$ of the frames should fall below the threshold.
Applying this threshold to the frame-level scores from speech samples generated by the TTS systems gives us all segments which are of same or lower quality than the lowest quality segments in the reference.
To remove spurious detections, we run a sliding window with a window length of $\SI{200}{ms}$ over the binary detections and remove any detections which are shorter than $\SI{100}{ms}$.
This approach results in 857 detected segments for \texttt{test-clean} (407 utterances), 2454 detected segments for F5-TTS (1481 utterances), and 911 detected segments for StyleTTS2 (555 utterances)\footnote{Samples: \url{https://go.upb.de/icassp26-sqa-detect}}.

\subsection{Analyzing the detected artefacts}
\begin{figure}[t]
    \centering
    \includegraphics[width=\linewidth]{%
        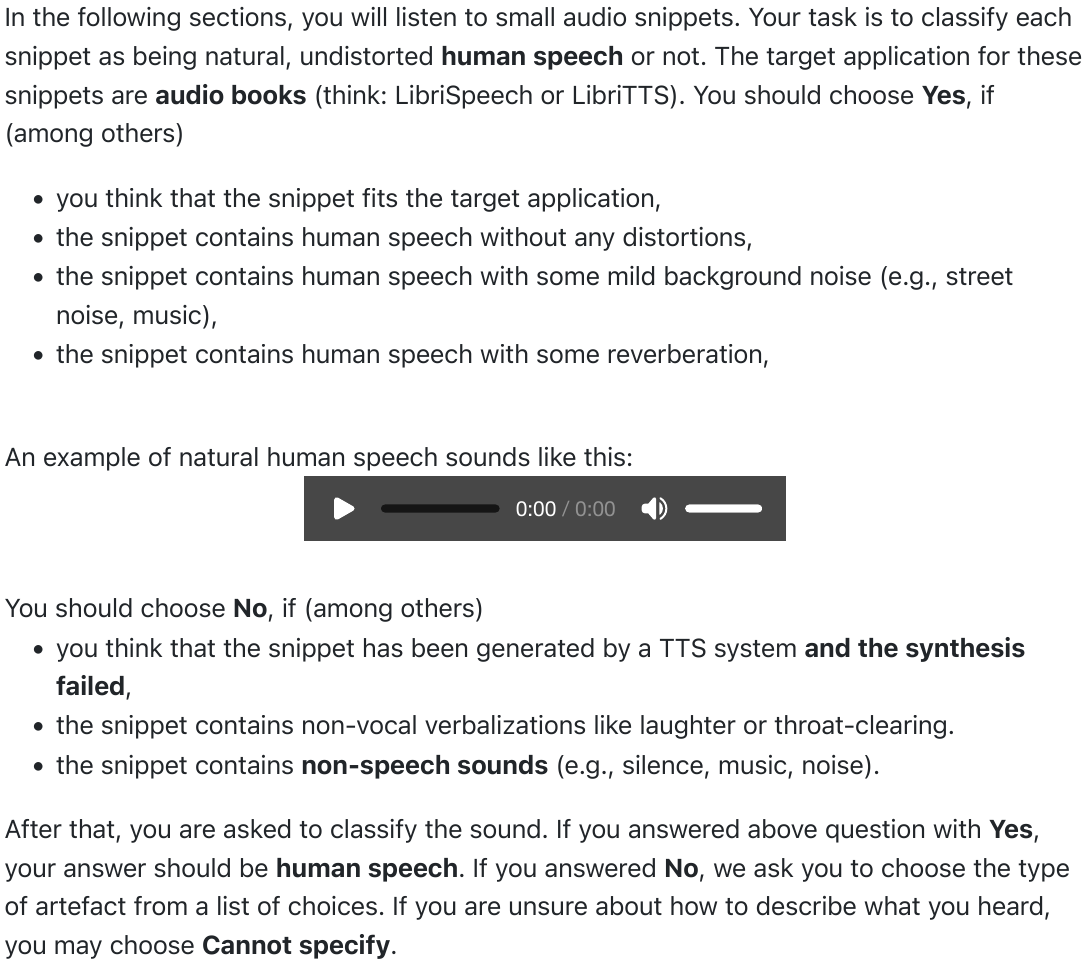%
    }
    \caption{Evaluation instructions for listeners.}
    \label{fig:instructions}
\end{figure}
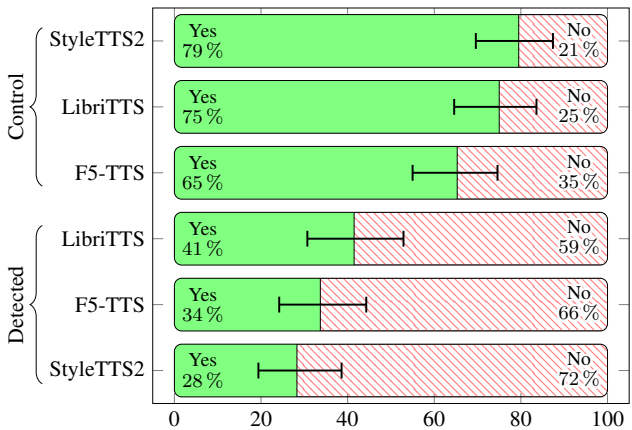
\begin{figure}[t]
    \centering
    \resizebox{\linewidth}{!}{
\begin{tikzpicture}[
    stackedbar/.style 2 args={
        rounded corners=3,
        path picture={
            \draw[black,rounded corners=0] (path picture bounding box.north east) rectangle (path picture bounding box.south west);
            \fill[green!50,rounded corners=0] (path picture bounding box.north west) rectangle ($(path picture bounding box.south west)!#1!(path picture bounding box.south east)$);
            \pgfmathtruncatemacro\ptrue{round(#1 * 100)}
            \node[anchor=west,align=center] at (path picture bounding box.west) {\footnotesize Yes \\[-.25em]\footnotesize \SI{\ptrue}{\percent}};
            \fill[pattern color=red!50,pattern=north west lines,rounded corners=0] ($(path picture bounding box.south west)!#1!(path picture bounding box.south east)$) rectangle (path picture bounding box.north east);
            \pgfmathtruncatemacro\pfalse{round(#2 * 100)}
            \node[anchor=east,align=center] at (path picture bounding box.east) {\footnotesize \contour{white}{No} \\[-.25em]\footnotesize \contour{white}{\SI{\pfalse}{\percent}}};
            \draw ($(path picture bounding box.south west)!#1!(path picture bounding box.south east)$) -- ($(path picture bounding box.north west)!#1!(path picture bounding box.north east)$);
        }
    }
]
  \begin{axis}[
    xmin=-.05, xmax=1.05,
    ymin=-.5, ymax=5.5,
    ytick={0,1,2,3,4,5},
    yticklabels={StyleTTS2,F5-TTS,LibriTTS,F5-TTS,LibriTTS,StyleTTS2},
    xtick={0,0.2,0.4,0.6,0.8,1},
    xticklabels={0,20,40,60,80,100},
    clip=false,
  ]
    \def\cap{0.12}
    \foreach \ntrue/\nfalse/\cilow/\cihigh [count=\i from 0] in {
      0.283/0.717/0.194/0.386,
      0.337/0.663/0.242/0.443,
      0.415/0.585/0.307/0.529,
      0.653/0.347/0.550/0.746,
      0.750/0.250/0.646/0.836,
      0.795/0.205/0.696/0.874
    }{
        \edef\temp{
            \noexpand\draw[stackedbar={\ntrue}{\nfalse}] (0, \i-.4) rectangle (1, \i+.4) {};
            \noexpand\draw[thick] (\cilow,\i) -- (\cihigh,\i);
            \noexpand\draw[thick] (\cilow,\i-\cap) -- (\cilow,\i+\cap);
            \noexpand\draw[thick] (\cihigh,\i-\cap) -- (\cihigh,\i+\cap);
        }
        \temp
    }

    \draw [decorate,decoration={brace,amplitude=5pt}] (yticklabel cs:.05) -- (yticklabel cs:.45) node[midway,rotate=90,yshift=1.2em]{Detected};
    \draw [decorate,decoration={brace,amplitude=5pt}] (yticklabel cs:.55) -- (yticklabel cs:.95) node[midway,rotate=90,yshift=1.2em]{Control};

  \end{axis}
\end{tikzpicture}
    }
    \caption{%
        Breakdown of \enquote{is human speech} answers for the evaluated speech segments.
        Black: 95\% confidence interval.
    }
    \label{fig:eval-human}
\end{figure}

To get insight about the types of detected artefacts, we conducted a listening test.
The goal of the listening test was to answer two questions:
1) What is the percentage of the detected segments that did \textbf{not} contain human-like speech fitting the target application, i.e., how good is the \emph{precision}?
2) Do the artefact types cluster by system?
\Cref{fig:instructions} shows the detailed set of instructions that were provided to the listeners.
For each system, we evaluated the $100$ lowest scored segments.
We observed that both systems sometimes failed to synthesize actual speech.
We detected and removed such failure cases if the synthesized speech was less than $\SI{1}{s}$ long or if a forced alignment failed or was not able to align more than one word.
To verify that our detection approach indeed detected the segments with the most detrimental quality, we added, for each detection set, control samples of the same amount, where a control sample is a random segment of 5 words from the full set of synthesized or clean (in case of \texttt{test-clean}) samples.
In total, $2\times3$ sets, each with 100 samples, had to be evaluated.
We recruited 9 speech technology experts, where each expert evaluated 60 samples (10 per set), which gave one rating per sample, leaving 60 samples without any rating.
We used \texttt{replikant}~\cite{lemaguer25_interspeech} to setup and run the experiment.

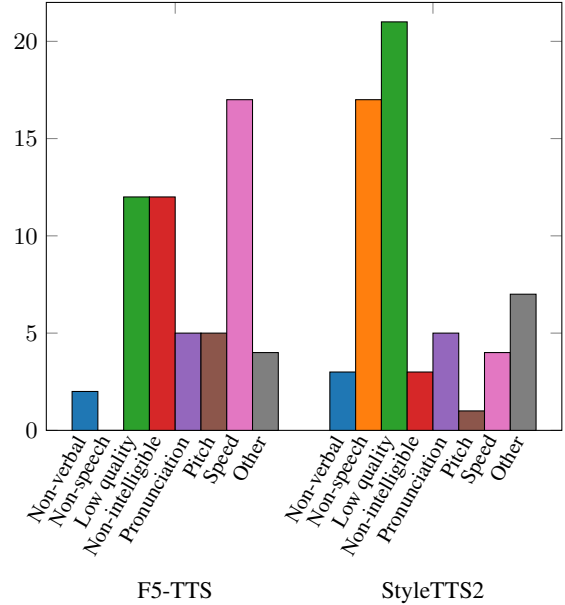
\begin{figure}[t]
    \centering

\definecolor{palette-0}{HTML}{1f77b4}
\definecolor{palette-1}{HTML}{ff7f0e}
\definecolor{palette-2}{HTML}{2ca02c}
\definecolor{palette-3}{HTML}{d62728}
\definecolor{palette-4}{HTML}{9467bd}
\definecolor{palette-5}{HTML}{8c564b}
\definecolor{palette-6}{HTML}{e377c2}
\definecolor{palette-7}{HTML}{7f7f7f}
\definecolor{palette-8}{HTML}{bcdb22}
\definecolor{palette-9}{HTML}{17becf}

\begin{tikzpicture}
    \begin{axis}[
        xtick={0,1},
        xticklabels={F5-TTS,StyleTTS2},
        xticklabel style={yshift=-6em},
        xmin=-.5, xmax=1.5,
        ymin=0,
        ymax=22,
        clip=false,
    ]
        \def\groupwidth{0.8}
        \def\barwidth{\groupwidth/8}   
        \foreach \system/\amount/\reason/\reasontext in {0/2.0/0/{Non-verbal},1/3.0/0/{Non-verbal},0/0/1/{Non-speech},1/17.0/1/{Non-speech},0/12.0/2/{Low quality},1/21.0/2/{Low quality},0/12.0/3/{Non-intelligible},1/3.0/3/{Non-intelligible},0/5.0/4/{Pronunciation},1/5.0/4/{Pronunciation},0/5.0/5/{Pitch},1/1.0/5/{Pitch},0/17.0/6/{Speed},1/4.0/6/{Speed},0/4.0/7/{Other},1/7.0/7/{Other}} {
            \edef\temp{
                \noexpand\draw[fill=palette-\reason] (\system+\reason*\barwidth-\groupwidth/2,0) rectangle (\system+\reason*\barwidth + \barwidth-\groupwidth/2,\amount);
                \noexpand\node[anchor=east,rotate=60,font=\noexpand\footnotesize] at (\system+\reason*\barwidth-\groupwidth/2 + 0.5*\barwidth,0) {\reasontext};
            }
            \temp
        }
    \end{axis}
\end{tikzpicture}
    \caption{%
        Absolute counts of identified artefact types by system.
    }
    \label{fig:eval-reasons}
\end{figure}

\Cref{fig:eval-human} shows a breakdown of the answers to the first question (\enquote{Does the audio snippet contain natural, undistorted human speech?}) by evaluation set.
All control sets receive a higher percentage of positive answers than the sets that contain speech segments from the quality detection.
The high percentage of positive answers for the StyleTTS2 control samples underlines its high quality and suitability for the target application.
However, it is also the system whose detections were most often classified as not befitting the target application.

We provide an overview about the kind of artefact types that were attributed to the detections by the listeners in~\Cref{fig:eval-reasons}.
Regarding StyleTTS2, the two most common types of artefacts were identified as \enquote{non-speech} and speech having \enquote{low quality}, where low quality means that there are severe quality degradations like low signal-to-noise ratio, clipping effects or buzzing sounds present in the audio.
The segments that were classified as \enquote{non-speech} are samples where StyleTTS2 failed to synthesize any speech.
These should have been filtered out by the filtering stage but where not detected.
Regarding F5-TTS, the most common type of artefact was \enquote{speed}, which always meant that the words were spoken too fast without any noticable pauses.
We assume that this stems from a misalignment between the target text and the reference sample for voice cloning, and the system seemingly tried to compress a long text into a short audio.



\section{CONCLUSION \& DISCUSSION}
In this contribution, we looked into the computation of frame-level scores for \emph{localizable} SQA.
We showed that the use of a consistency loss in the training of SQA models improves the localization of segments of poor speech quality in artificially generated speech.
We believe that this is a step towards better interpretabiliy of automatic quality predictions, as this can localize and identify error patterns future improvements of a TTS system should focus on.

We noticed that the types of artefacts the SQA model reacted sensitive to depend on the training data.
In an informal listening test, we found that the model assigned low scores to non-verbal vocalizations like laughter or throat clearing, likely because these were rarely seen during training.
Depending on the target application, this may produce undesired false positives.
Future work should therefore look into the alignment of the SQA model with the target application.


\bibliographystyle{IEEEbib}
\bibliography{strings,refs}

\end{document}